\documentclass{iaus}
\usepackage{graphicx}

\title[The magnetic field in NGC\,253] 
{The magnetic field structure in NGC\,253 in presence of a galactic wind}

\author[V.\,Heesen et al.]   
{V.\,Heesen$^1$, M.\,Krause$^2$, R.\,Beck$^2$, and R.-J.\,Dettmar$^1$}
\affiliation{$^1$Astronomisches Institut der Ruhr-Universit\"at
  Bochum, 44780 Bochum, Germany \break email:
  heesen@astro.rub.de, dettmar@astro.rub.de\\[\affilskip]
$^2$Max-Planck-Institut f\"ur Radioastronomie, 53121 Bonn, Germany
\break email: rbeck@mpifr-bonn.mpg.de, mkrause@mpifr-bonn.mpg.de}

\pubyear{2009}
\volume{259}  
\pagerange{119--126}
\date{?? and in revised form ??}
\jname{Proceedings Title IAU Symposium}
\editors{K.G.\,Strassmeier, A.G.\,Kosovichev, \& J.\,Beckmann }
\begin{document}

\maketitle

\begin{abstract}
We present radio continuum polarimetry observations of the nearby
edge-on galaxy NGC\,253 which possesses a very bright radio halo. Using
the vertical synchrotron emission profiles and the lifetimes of
cosmic-ray electrons, we determined the cosmic-ray bulk speed
as $300\pm30\,\rm km\,s^{-1}$, indicating the presence of a galactic
wind in this galaxy. The large-scale magnetic field was decomposed
into a toroidal axisymmetric component in the disk and a poloidal
component in the halo. The poloidal component shows a prominent
X-shaped magnetic field structure centered on the nucleus, similar
to the magnetic field observed in other edge-on galaxies. Faraday
rotation measures indicate that the poloidal field has an odd parity
(antisymmetric). NGC\,253 offers the possibility to compare the
magnetic field structure with models of galactic dynamos and/or
galactic wind flows. \keywords{galaxies: halos, galaxies: individual
(NGC253), galaxies: magnetic fields, galaxies: spiral, radio
continuum: galaxies}
\end{abstract}

\firstsection 
\section{Introduction}
Gaseous halos are a common property of star forming galaxies. The
abundance of hot coronal gas in the halo requires heating from the
star forming disk and a transport of energy from the disk into the
halo. This energy transport has been discussed as galactic fountains
or chimneys \cite[(Field et al.\ 1969, Norman \& Ikeuchi
  1989)]{field_69a, norman_89a}. The relativistic cosmic-ray gas has
an adiabatic index of 4/3 compared to 5/3 of the (single atomic) hot
gas; this results in a larger pressure scaleheight of the cosmic-ray
gas. The magnetic field has an even larger pressure scaleheight
\cite[(Beck 2009)]{beck_09}. The pressure contributions of the cosmic
rays and the magnetic field are thus dominating in the halo and can
accelerate the outflow. This led to the model of a cosmic-ray driven
wind that causes acceleration of the flow accelerates further away
from the disk \cite[(Breitschwerdt et al.\ 1991)]{breitschwerdt_91a}.
The existence of radio halos around many nearby galaxies shows the
presence of cosmic rays and magnetic fields. Cosmic rays spiral around
magnetic field lines and follow them from the disk into the
halo. Hence, a
vertical component of the large-scale magnetic field can enhance
the vertical cosmic-ray transport. How to generate vertical fields is
still an open question. \cite[Parker (1992)]{parker_92} suggested that
field lines start to overlap and reconnect to form ``open'' field
lines that connect the disk with the halo. Another possibility is the
mean field $\alpha\Omega$-dynamo, which creates dipolar or quadrupolar
field configurations.  Furthermore, the galactic wind may shape the
magnetic field and align with it. This may particularly hold for
\emph{superwinds}, which originate in a region of very active star
formation (starburst).
As shown by X-ray observations, NGC\,253 has a superwind and is an
ideal object to study the interaction between the superwind and the
magnetic field. There is a nuclear plume of H$\alpha$ and X-ray
emitting gas \cite[(Strickland et al.\ 2000, Bauer et
  al.\ 2007)]{strickland_00a,bauer_07a} where the outflow velocity was
directly measured by spectroscopy as $390\,\rm km\,s^{-1}$
\cite[(Schulz \& Wegner 1992)]{bauer_07a,schulz_92a}. The connection
between the heated gas in the halo, which extends far from the disk,
and the the superwind is likely as suggested by numerical simulations
of a centrally driven wind \cite[(Suchkov et
  al.\ 1994)]{suchkov_94a}. A detailed study of this galaxy is
possible because it is nearby ($D=3.94\,\rm Mpc$, \cite[Karachentsev
  et al.\ 2003]{karachentsev_03a}, $30''=600\,\rm pc$). With an
inclination angle of $78^\circ$ the galaxy's disk is only mildly
edge-on. But it contains a very bright nucleus which complicates the
data reduction of radio continuum observations.
\section{Observations}
Our new results are based on a deep mosaic with the VLA at $\lambda
6.2\,\rm cm$. This wavelength is optimal to study magnetic fields in
galactic halos, because the polarization is not too much affected by
Faraday rotation and depolarization. On the other hand, the
contribution from the thermal emission is still small ($<10\%$) and
the vertical extent is large. The largest angular scale that can be
well imaged with the VLA at $\lambda 6.2\,\rm cm$ is $5'$. We used a
$\lambda 6.3\,\rm cm$ map observed with the 100-m
Effelsberg telescope to fill in the missing zero-spacing flux.
Additionally we used VLA maps at $\lambda 20\,\rm cm$ \cite[(Carilli
  et al.\ 1992)]{carilli_92a} and at $\lambda 90\,\rm cm$
\cite[(Carilli 1996)]{carilli_96a}. In order to correct the magnetic
field for Faraday rotation we observed the galaxy with the 100-m
Effelsberg telescope at $\lambda 3.6\,\rm cm$. A detailed description
of the observations and the data reduction can be found in
\cite{heesen_08a}.
\section{Morphology}
\begin{figure}[tb]
\begin{minipage}[b]{0.66\textwidth}
\resizebox{\hsize}{!}{\includegraphics[bb=35 160 565 665,clip=]{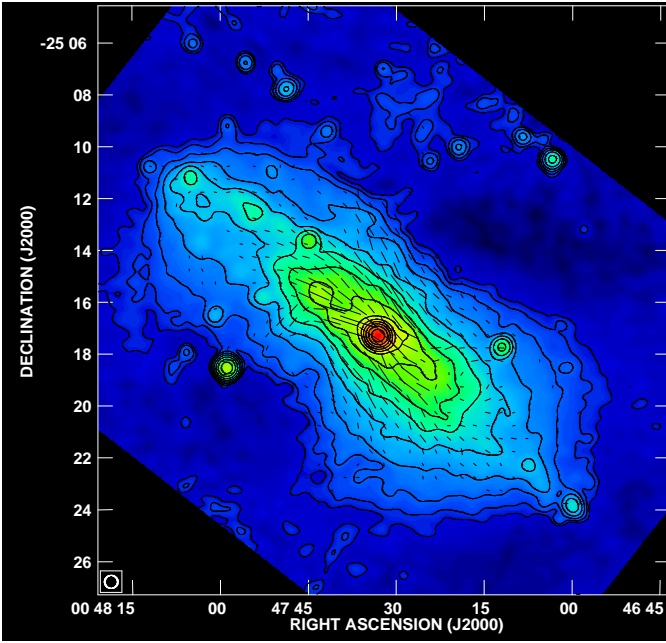}}
\end{minipage}
\hfill
\begin{minipage}[b]{0.33\textwidth}
\caption{Total power radio continuum at $\lambda 6.2\,{\rm cm}$
  obtained from the combined VLA + Effelsberg observations with $30''$
  resolution. Contours are at 3, 6, 12, 24, 48, 96, 192, 384, 768,
  1536, 3077, 6144, 12288, and 24576 $\times$ the rms noise of $30\,{\rm\mu
    Jy/beam}$. The overlaid vectors indicate the orientation of the
  Faraday corrected regular magnetic field. A vector length of $1''$ is
  equivalent to $12.5\,{\mu\rm Jy/beam}$  polarized intensity.}
\label{fig:n253cm6ve_tp_b30}
\end{minipage}
\end{figure}

\begin{figure}[tb]
\begin{minipage}[b]{0.66\textwidth}
\resizebox{\hsize}{!}{\includegraphics[bb=5 60 550 515,clip=]{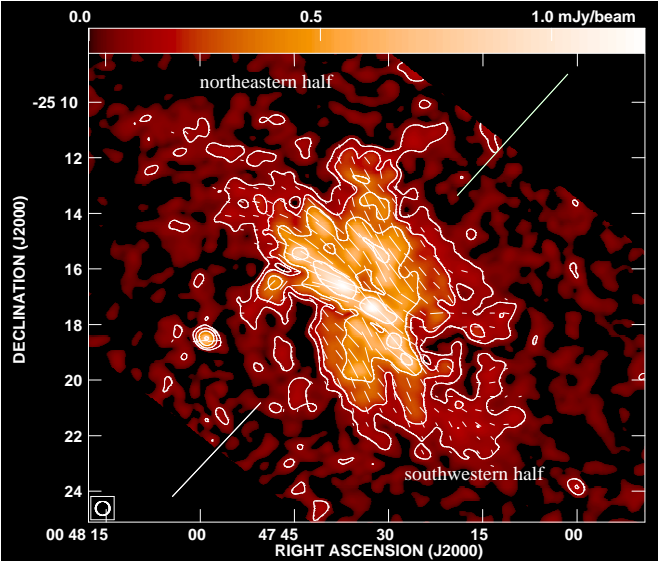}}
\end{minipage}
\hfill
\begin{minipage}[b]{0.33\textwidth}
\caption{Polarized intensity obtained from the combined VLA +
  Effelsberg observations at $\lambda 6.2\,{\rm cm}$ with $30''$
  resolution. Contours are at 3, 6, 12, and 24 $\times$ the rms noise
  level of $30\,{\mu\rm Jy/beam}$. The overlaid vectors indicate the
  orientation of the Faraday corrected regular magnetic field. A vector length
  of $1''$ is equivalent to $12.5\,{\mu\rm Jy/beam}$ polarized
  intensity.}
\label{fig:n253cm6ve_piaf_b30}
\end{minipage}
\end{figure}

In Fig.\,\ref{fig:n253cm6ve_tp_b30} we present the $\lambda 6.2\,\rm
cm$ total power radio continuum emission from the combined VLA and
Effelsberg observations. The vectors show the Faraday corrected
orientation of the large-scale magnetic field. The emission shows a
prominent dumbbell shaped radio halo. The vertical extent is smallest
near the center and increases further out in the disk. A detailed
analysis of the vertical emission profiles shows that the scaleheights
are smallest near the center.
The distribution of the polarized emission shown in
Fig.\,\ref{fig:n253cm6ve_piaf_b30} is more concentrated to the inner
disk than the total power emission. It is extending into the halo
particularly in the inner disk. The orientation of the magnetic field
is disk-parallel in the inner part of the disk but turns away from
the disk in the outer part. The vertical component even dominates at
the ``radio spur'' ($\rm
R.A.=00^h47^m50^s$, $\rm Dec.=-25^\circ17^m00^s$), where the magnetic
field orientation is almost perpendicular to the disk.
\section{Cosmic-ray bulk speed}
\begin{figure}[tb]
  \resizebox{\hsize}{!}{
    \LARGE\input{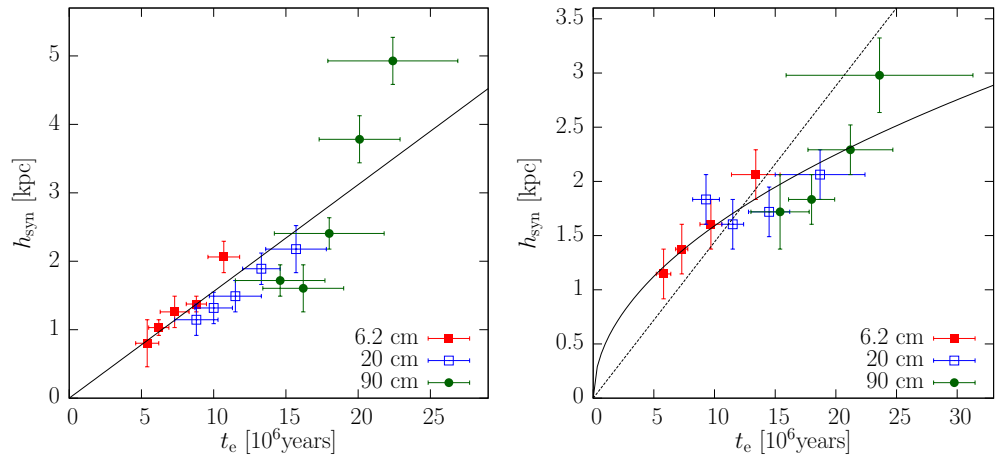}}
\caption{Scaleheight $h_{\rm syn}$ of the thick radio disk as a
  function of the electron lifetime $t_{\rm e}$ in the northeastern
  halo (left) and in the southwestern halo (right). The linear fit is
  the theoretical expectation for a convective cosmic-ray transport
  with a constant bulk speed. The fit $h_{\rm e}\propto\sqrt{t_{\rm e}}$ is the
  theoretical expectation for a diffusive cosmic-ray transport.}
\label{fig:h_sl_col}
\end{figure}

The dumbbell shaped radio halo shows that the vertical extension is
smallest in the inner part of the disk where the magnetic field
strength is highest. Synchrotron losses
are the dominant energy loss of the cosmic-ray electrons
\cite[(Heesen et al.\ 2009a)]{heesen_09a}. The synchrotron lifetime
\begin{equation}
t_{\rm syn}=\frac{8.352\times 10^9\,{\rm yr}}{(\nu / 16.1\,\rm
  MHz)^{1/2}(B_\perp / \mu{\rm G})^{3/2}}
\end{equation}
depends on the total magnetic field strength $B_\perp$ perpendicular
to the line-of-sight and on the observation frequency $\nu$.  Thus, it
is smallest in the central part of the disk. To calculate the electron
lifetime we have to add the timescale $t_{\rm a}$ of the adiabatic losses:
\begin{equation}
\frac{1}{t_{\rm e}}=\frac{1}{t_{\rm a}}+\frac{1}{t_{\rm syn}}.
\end{equation}
Figure~\ref{fig:h_sl_col} shows the scaleheight of the synchrotron
emission as a function of the electron lifetime. The linear
dependence, particularly in the northeastern halo of the galaxy
(c.f.\,Fig.\,\ref{fig:n253cm6ve_piaf_b30}), suggests a definition of an
average cosmic-ray bulk speed
\begin{equation}
{\rm v}=\frac{h_{\rm e}}{t_{\rm e}},
\end{equation}
where $h_{\rm e}$ is the electron scaleheight (twice the synchrotron
scaleheight). In the northeastern
halo, we find a cosmic-ray bulk speed of $300\pm30\,\rm km\,s^{-1}$.
Because the cosmic-ray bulk speed does not depend on the electron
energy, the cosmic-ray transport is convective. In the southwestern
halo, the scaleheight can be better described by $h_{\rm
  e}\propto \sqrt{t_{\rm e}}$. The cosmic-ray bulk speed increases
with the electron energy, which indicates a
larger role of diffusion in this halo part. In case of pure diffusion
we have a diffusion coefficient of $\kappa=2.0\pm0.2\times10^{29}\,\rm
cm^2\,s^{-1}$.
Because the cosmic-ray electrons lose their energy via synchrotron
radiation, their energy spectral index increases during their
lifetime. This is visible as a steepening of the radio spectral
index in the halo. We detected an almost linear dependence of the
radio spectral index on the distance from the disk. From this we
derived $170\,\rm km\,s^{-1}$ as a lower limit for the average
cosmic-ray bulk speed.
\section{Structure of the magnetic field}
\begin{figure}[tb]
\begin{minipage}[b]{0.66\textwidth}
\resizebox{\hsize}{!}{\includegraphics[bb=30 170 570 635,clip=]{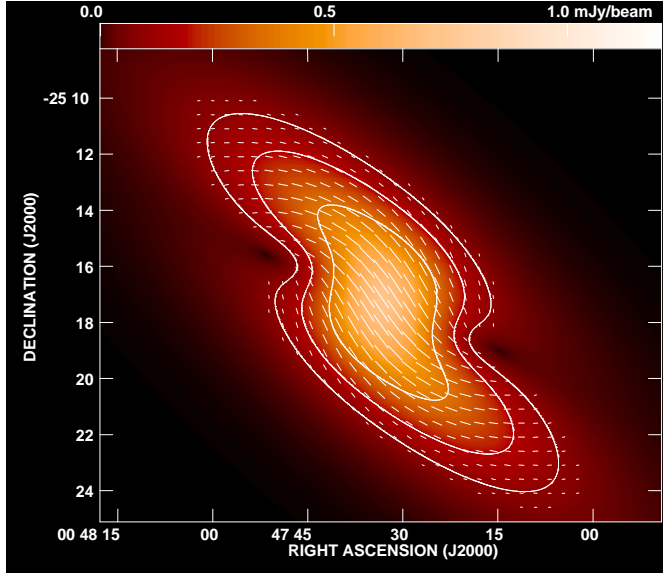}}
\end{minipage}
\hfill
\begin{minipage}[b]{0.33\textwidth}
\caption{Polarized emission and magnetic field orientation expected
  for an axisymmetric spiral model for the toroidal magnetic field at
  $180''$ resolution. Contours and vectors as in
  Fig.\,\ref{fig:n253cm6ve_piaf_b30}.}
\label{fig:pa25_g700_pia_ps}
\end{minipage}
\end{figure}

\begin{figure}[tb]
\begin{minipage}[b]{0.66\textwidth}
\resizebox{\hsize}{!}{\includegraphics[bb=5 60 535 562,clip=]{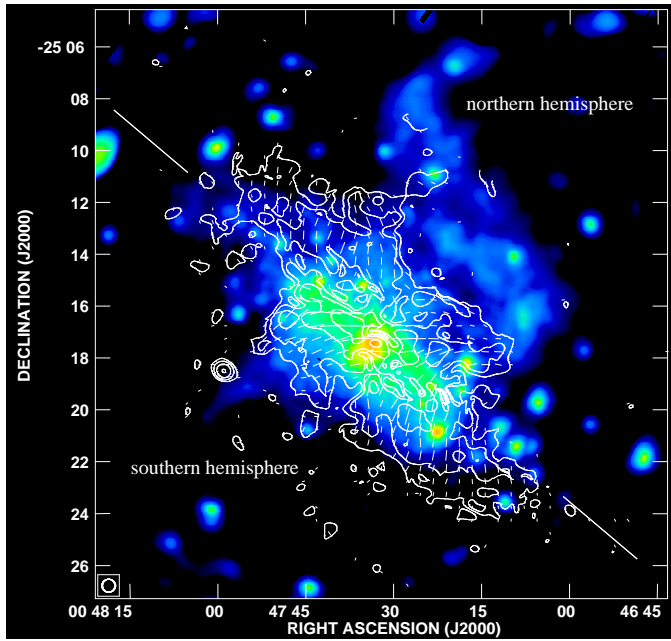}}
\end{minipage}
\hfill
\begin{minipage}[b]{0.33\textwidth}
\caption{Polarized emission and magnetic field orientation of the
  poloidal magnetic field overlaid onto diffuse X-ray
  emission. Contours and vectors as in
  Fig.\,\ref{fig:n253cm6ve_piaf_b30}. The X-ray map is from XMM
  observations in the energy band $0.5-1.0\,\rm keV$ (\cite[Bauer et
    al.\ 2008]{bauer_08a}).}
\label{n253_pol_XMM2_col}
\end{minipage}
\end{figure}

Our model of an axisymmetric spiral, disk-parallel \emph{toroidal}
magnetic field (Fig.\,\ref{fig:pa25_g700_pia_ps}) represents the
observations well. We chose the inclination angle of the optical disk
and a best-fit spiral pitch angle of 25$^\circ$, which is similar to
the pitch angle of 20$^\circ$ of the optical spiral arms. At the
locations where the toroidal magnetic field is weak we also observe a
vertical \emph{poloidal} field
(Fig.\,\ref{fig:n253cm6ve_piaf_b30}). We subtracted the toroidal field
model from the observations and obtained the poloidal magnetic field
shown in Fig.\,\ref{n253_pol_XMM2_col}. It has a prominent X-shaped
structure centered on the nucleus.
The structure of the poloidal magnetic field seems to be connected to
the distribution of the heated gas in the halo. This gas forms a
horn-like structure with lobes that are thought to be the walls of a
superbubble filled with dilute hot gas. This can be seen in
Fig.\,\ref{n253_pol_XMM2_col} where the sensitive XMM observations
show X-ray emitting gas also inside the bubble. The similarity of the
structure of the poloidal magnetic field to that of the superbubble
and their alignment can be explained by an interaction between the
superwind, driven by the nuclear starburst, and the surrounding medium
transported by the disk wind, possibly enhanced by shock waves of the
expanding superbubble.
The difference between the polarization angles at $\lambda\lambda$
$6.2\,\rm cm$ and $3.6\,\rm cm$ is due to Faraday rotation. This
provides information about the line-of-sight component of the
large-scale magnetic field. \cite{krause_89a} showed that the
azimuthal variation of rotation measure (RM) can be used to determine
the contributions from individual dynamo modes. In NGC\,253, the RM is
the superposition of the toroidal magnetic field in the disk and the
poloidal magnetic field in the halo. Subtracting the model for the
disk magnetic field leaves the RM distribution of the poloidal
magnetic field. It has the maximum amplitude along the minor axis. We
propose a conical configuration for the poloidal magnetic field with
an opening angle of $66^\circ$ of odd symmetry: the field direction
points towards the disk in the southern hemisphere
(c.f.\,\ref{n253_pol_XMM2_col}) and away from the disk in the
northern hemisphere \cite[(Heesen et al.\ 2009)]{heesen_09b}. The field
lines are aligned with the lobes of hot gas in the halo .
\section{Conclusions}
NGC\,253 possesses a bright radio halo. The scaleheight of the
synchrotron emission depends on the lifetime of the cosmic-ray
electrons. This requires a vertical cosmic-ray transport from the disk
into the halo. The disk wind has an average velocity of $300\pm30\,\rm
km\,s^{-1}$ and is surprisingly constant over the extent of the
disk. The disk wind transports material from the disk into the
halo. If this material is the origin of the luminous, heated gas in
the halo, it can explain the asymmetry between the northeastern and
southwestern halo, as the cosmic-ray transport is much more
efficient in the convective northeastern halo than in the diffusive
southwestern halo.
In the disk the magnetic field is parallel to the disk. A model of an
axisymmetric spiral disk field can explain the observed
asymmetries in polarization by a geometrical effect. After subtracting
the disk field model from the observations we obtain the poloidal
magnetic field in the halo. This field has a prominent X-shaped
structure that is also seen in several other edge-on galaxies. For the
first time the parity of the large-scale field in any galaxy could be
determined, which is odd (antisymmetric) in NGC\,253. The X-shape may
be explained by interaction of the halo gas transported by the disk
wind with the superwind from the starburst center. However, several
galaxies without a superwind also show X-shaped halo fields
\cite[(Krause 2008)]{krause_08a}, which needs to be further
investigated.

\end{document}